\documentclass[amsmath,prd,groupedaddress,
nobibnotes,nofootinbib,floatfix]  
{revtex4}    
\usepackage{graphicx}
\usepackage{bm}

\def\ba{\begin{eqnarray}}
\def\ea{\end{eqnarray}}
\def\be{\begin{equation}}
\def\ee{\end{equation}}
\def\({\left(}
\def\){\right)}
\def\[{\left[}
\def\]{\right]}
\def\<{\left<}
\def\>{\right>}

\newcommand{\rar}{\rightarrow}

\newcommand{\labeq}[1] {\label{eq:#1}}
\newcommand{\refeq}[1] {(\ref{eq:#1})}

\newcommand{\labsec}[1] {\label{sec:#1}}
\newcommand{\refsec}[1] {Sec.~\ref{sec:#1}}
\newcommand{\refapp}[1] {Appendix \ref{sec:#1}}

\newcommand{\wmap}{{WMAP}}

\newcommand{\cmb}{{CMB}}

\newcommand{\ilc}{{ILC}}
\newcommand{\ica}{{ICA}}

\newcommand{\mbS}{\ensuremath{\mathbf{S}}}
\newcommand{\mbU}{\ensuremath{\mathbf{U}}}
\newcommand{\mbx}{\ensuremath{\mathbf{x}}}

\newcommand{\mbX}{\ensuremath{\mathbf{X}}}
\newcommand{\mba}{\ensuremath{\mathbf{a}}}
\newcommand{\mbA}{\ensuremath{\mathbf{A}}}

\newcommand{\mbM}{\ensuremath{\mathbf{M}}}
\newcommand{\mbN}{\ensuremath{\mathbf{N}}}
\newcommand{\mbG}{\ensuremath{\mathbf{G}}}

\newcommand{\mbAm}{\ensuremath{\mathbf{A}^{-1}}}
\newcommand{\mbAmt}{\ensuremath{\mathbf{A}^{-\text{T}}}}
\newcommand{\mbXt}{\ensuremath{\mathbf{X}^\text{T}}}
\newcommand{\mbNm}{\ensuremath{\mathbf{N}^{-1}}}
\newcommand{\mbGm}{\ensuremath{\mathbf{G}^{-1}}}
\newcommand{\mbc}{\ensuremath{\mathbf{c}}}
\newcommand{\mbcm}{\ensuremath{\mathbf{c}^{-1}}}

\newcommand{\mbC}{\ensuremath{\mathbf{C}}}
\newcommand{\mbCm}{\ensuremath{\mathbf{C}^{-1}}}
\newcommand{\mbi}{\ensuremath{\mathbf{i}}}

\newcommand{\mbw}{\ensuremath{\mathbf{w}}}
\newcommand{\mbwt}{\ensuremath{\mathbf{w}^\text{T}}}
\newcommand{\dmbw}{\ensuremath{\delta\mathbf{w}}}
\newcommand{\dmbwt}{\ensuremath{\delta\mathbf{w}^\text{T}}}
\newcommand{\mbam}{\ensuremath{\mathbf{a}^{-1}}}
\newcommand{\mbamt}{\ensuremath{\mathbf{a}^{-\text{T}}}}
\newcommand{\mbu}{\ensuremath{\mathbf{u}}}
\newcommand{\mbut}{\ensuremath{\mathbf{u}^\text{T}}}

\newcommand{\mbe}{\ensuremath{\mathbf{e}}}
\newcommand{\mbet}{\ensuremath{\mathbf{e}^\text{T}}}
\newcommand{\mbD}{\ensuremath{\mathbf{D}}}
\newcommand{\mbF}{\ensuremath{\mathbf{F}}}
\newcommand{\mbDm}{\ensuremath{\mathbf{D}^{-1}}}
\newcommand{\mbFm}{\ensuremath{\mathbf{F}^{-1}}}
\newcommand{\mbQ}{\ensuremath{\mathbf{Q}}}

\newcommand{\mbQt}{\ensuremath{\mathbf{Q}^\text{T}}}

\newcommand{\mbzero}{\ensuremath{\mathbf{0}}}

\newcommand{\mbP}{\ensuremath{\mathbf{P}}}
\newcommand{\mbPt}{\ensuremath{\mathbf{P}^\text{T}}}

\newcommand{\vcw}{\ensuremath{\text{vec}(\mbw)}}
\newcommand{\vcwt}{\ensuremath{\text{vec}(\mbw)^\text{T}}}

\newcommand{\vci}{\ensuremath{\text{vec}(\mbi)}}
\newcommand{\vcit}{\ensuremath{\text{vec}(\mbi)^\text{T}}}

\newcommand{\tr}{\text{tr}\:}
\newcommand{\subsp}{_\text{s.p.}}

\newcommand{\mbn}{\ensuremath{\mathbf{n}}}

\newcommand{\hatx}{\ensuremath{\widehat{\mathbf{x}}}}
\newcommand{\hatxm}{\ensuremath{\widehat{\mathbf{x}}^{-1}}}

\newcommand{\hatR}{\ensuremath{\widehat{\mathbf{R}}}}
\newcommand{\hatRm}{\ensuremath{\widehat{\mathbf{R}}^{-1}}}

\newcommand{\cl}{\ensuremath{C_l}}

\newcommand{\nelem}{\ensuremath{{N_\text{elem}}}}

\newcommand{\ncmb}{\ensuremath{{n_\textsc{cmb}}}}
\newcommand{\nproj}{\ensuremath{{n_\text{proj}}}}

\newcommand{\dtsq}{\ensuremath{\Delta T^2}}
\newcommand{\txtfm}{\text{f.g.\ model}}
\newcommand{\txtcm}{\text{\cmb\ model}}

\newcommand{\mbSc}{\ensuremath{\mathbf{S}_{\textsc{cmb}}}}
\newcommand{\mbSct}{\ensuremath{\mathbf{S}_{\textsc{cmb}}^{\text{T}}}}

\newcommand{\lmax}{\ensuremath{l_\text{max}}}

\begin{document}

\title{Prescription for Cosmic Information Extraction from Multiple Sky Maps}

\date{\today}
\author{Steven Gratton}
\email{stg20@cam.ac.uk}
\affiliation{Insitute of Astronomy, Madingley Road, Cambridge, CB3 0HA, UK} 

\begin{abstract}

This paper presents a prescription for distilling the information
contained in the cosmic microwave background radiation from
multiple sky maps that also contain both instrument noise
and foreground contaminants.  
The prescription is well-suited for
cosmological parameter estimation and accounts for uncertainties
in the cosmic microwave background extraction scheme.  The
technique is computationally viable at low resolution and may be
considered a natural and significant generalization of the ``Internal
Linear Combination'' approach to foreground removal.  An important
potential application is the analysis of the multi-frequency
temperature and polarization  
data from the forthcoming Planck satellite. 

\end{abstract}

\maketitle

\section{\labsec{intro}Introduction}

The detection and subsequent investigation of the cosmic microwave background
(\cmb) radiation over the past four decades has been essential to our current
understanding of the universe.  Initially providing evidence for an early hot
dense radiation-dominated phase to the
universe~\cite{1965ApJ...142..419P,1965ApJ...142..414D}, the  
\cmb's spatial distribution is now studied in 
precise detail (see \cite{Hinshaw:2008kr,Gold:2008kp,
  Nolta:2008ih,Dunkley:2008ie,Wright:2008ib, 
Hill:2008hx, Komatsu:2008hk} for the latest analysis of the \wmap\
  satellite data by the \wmap\ science team)
for clues both about the state of the universe at the beginning
of the radiation era and the universe's composition, structure and subsequent
evolution.  As cosmologists we have been exceedingly fortunate that, from our
vantage point of the earth, the galaxy (away from its plane) is both
sufficiently transparent and sufficiently lacking in emission for us to be able to readily
measure the intensity of the cosmic signal shining through.  The cosmic signal
is also 
moderately polarized and this carries important additional cosmological
information, in particular about the presence of gravity waves in the universe
(which could be generated by say a high energy inflationary phase before the
radiation era) and about the reionization of the universe.  Unfortunately,
with the signal being weaker and the polarization of the galactic emission
being less understood than its intensity, both the detection and analysis of
the polarization part of the signal are much more challenging.  In
addition, there
is also extragalactic contamination, such as that from point sources.
This paper
presents a technique to extract the cosmological information out of multiple
sky maps, including polarization 
ones, in the presence of both instrument noise and foregrounds.  The
technique rests upon the assumption that the foregrounds do not have
the same frequency response as the \cmb\ and requires that the sky
maps probe linearly independent parts of the spectrum of the sky
signal.           

The prescription presented here relates to and builds upon a number of
approaches already in the literature.  The COBE team considered three
approaches to separating galactic emission from the cosmic signal, one
involving 
modelling known emissions using non-\cmb\ data, another involving
fitting the maps to functions of given spectral index, and a final one
involving linearly combining their multifrequency maps to cancel the
dominant galactic emission~\cite{1992ApJ...396L...7B}.  The \wmap\ team
primarily use a template subtraction procedure to mitigate the effects
of foreground contaminants in power spectrum estimation, and study the 
foregrounds themselves via maximum entropy and most recently Markov
Chain Monte Carlo parametric methods.
However, 
initially for 
visualization purposes but later also for analysis they additionally 
introduced the ``Internal Linear Combination'' (\ilc) scheme, forming a
linear combination of their sky maps and choosing the weights to
minimize the variance between the maps whilst being constrained to
preserve unit response to the \cmb\
\cite{Bennett:2003ca,Hinshaw:2006ia,Gold:2008kp}.  An harmonic
mode-by-mode equivalent of \ilc\ was presented
in~\cite{1996MNRAS.281.1297T} and applied to the \wmap\ data in
\cite{Tegmark:2003ve}.  An alternative harmonic-based generalization
of the \ilc\ technique was recently presented in \cite{Kim:2008zh}. 
The ``Independent Component Analysis'' (\ica) signal
processing technique (see~\cite{hyvarinen}), which attempts to use
non-gaussianity of the one-pixel distribution to separate data into
independent signals, has been 
tested for cosmological uses on the COBE, BEAST and \wmap\
data~\cite{Maino:2001vz, Maino:2003as,Donzelli:2005is, Maino:2006pq}.
A possible weakness of 
\ica\ to \cmb\ extraction is that the \cmb\ is believed to be very
close to gaussian and so can only emerge as what is left behind after
the non-gaussian foregrounds have been removed.  The related
``Correlated Component Analysis'' idea uses pixel-pixel cross
correlations instead of non-quadratic single-pixel statistics to
separate signals and has also been applied to the \wmap\
data~\cite{Bonaldi:2006qw,Bonaldi:2007mf}.  A modification of \ica\ that forces it to
take into account the black body nature of the \cmb\ (a key feature of
the approach described here) was recently presented
in~\cite{Vio:2008kw}. 
The ``spectral matching'' approach
of~\cite{Delabrouille:2002kz,Patanchon:2004kj} 
shares many similarities to that presented here, and a recent
paper~\cite{Cardoso:2008qt} presents an ``additive component''-based
separation technique. \cite{Eriksen:2005dr} suggests fitting model
parameters at low resolution and then using these to solve for high
resolution maps; see also the very recent work~\cite{Stompor:2008sf}
for more on parametric component separation.
A Gibbs sampling based approach to component 
separation and \cmb\ power spectrum estimation was presented
in~\cite{Eriksen:2007mx} and applied to the \wmap\ data
in~\cite{Eriksen:2007mp}.  The \wmap\ team also test this approach
in~\cite{Dunkley:2008ie}.    

Unlike many of the above papers, this work focusses on likelihood
estimation for cosmological models as opposed to \cmb\ sky map production.
This requires a quantification of the uncertainties related to the
\cmb\ extraction (as also stressed by Ref.~\cite{Eriksen:2006pn}).  Of
course this is a somewhat ill-defined problem 
seeing as one 
does not know precisely what the foregrounds are.  By putting relatively
weak priors on the foregrounds though one might hope that such errors
are being estimated if anything conservatively.  Our prescription
allows one to naturally incorporate non-\cmb\ datasets and the
information they contain about the foregrounds into the analysis.
With the immiment 
launch of the Planck satellite~\cite{unknown:2006uk} 
and the significant 
new information on the polarization of the \cmb\ that it should
deliver, this work is also notable in treating all Stokes parameters
of the \cmb\ in a unified manner.  Numerical testing of the scheme and
application to existing \wmap\ data are underway.

\section{\labsec{priors}Data and Priors}

Our starting point shall be a collection of $n$ sky maps.  Any that are
usefully described in terms of a temperature will be assumed to be in
thermodynamic temperature units.  Note from the start that these maps
do not all have to be ``\cmb'' maps; other data sets (e.g.\ radio
surveys, starlight polarization maps, point source maps, the \wmap\
``spurious signal'' maps) can be included in
the analysis in a unified manner and might be useful if they have
physical correlations with contaminants in the \cmb\ channels.  

We assume the sky map is discretized into elements, typically pixels
or spherical 
harmonic coefficients up to some \lmax, but perhaps say wavelet coefficients.
For each element $i$ we have the  
associated measurements for each sky map.   We can stack these measurements into a vector $\mbX(i)$.  These 
vectors can be further stacked into a big vector \mbX\ (the entire data
set).  We'll assume we can estimate or
calculate the inverse noise covariance matrix \mbNm\  for \mbX.

Next, let us assume a linear relation between \mbX\ and some underlying
``signals'' \mbS, i.e.\
\ba
\mbX=\mbA \mbS + \mbM,
\ea
where \mbM\ is the noise realization.  Some of the signals will of
course be the \cmb, and the others will be the foregrounds.  These
``effective'' foreground 
signals don't necessarily have to be thought of as physical processes, just as
unwanted contaminants.  We
shall not
assume that these foregrounds are uncorrelated with each other, but
they will, however, almost by definition be taken to independent
of the \cmb.  To extract cosmological information,
we shall work towards a probability distribution for that part of
\mbS\ associated with the \cmb\ sky.

A key assumption shall be that
the \cmb\ is a blackbody.  As a consequence, the ``foreground''
signals will implicitly be assumed to be linearly independent of the
\cmb\ in frequency space.     

We'll typically assume that the ``mixing matrix'' \mbA\ is block-diagonal in pixel space, meaning that
measurements in a given direction only depend on the signals in that
direction (and we are assuming that any beam convolution effects have already been
accounted for).  In this case, we can write:
\ba
\mbX (i) =\mbA(ii) \mbS(i) + \mbM(i).
\labeq{locallink}
\ea 
(Here and onwards, a matrix followed by element indices denote the submatrix
appropriate to the indices of the original matrix.)  
$\mbS(i)$ encodes the strength of the signals, and $\mbA(ii)$ controls how much
each signal affects each instrument channel.  The degeneracy between the amplitude of a column of
$\mbA(ii)$ and the normalization of a given foreground signal is in principle
handled by assumed priors, and the discrete symmetry of column
interchange of $\mbA(ii)$ leads to an uniform multiplicative
overcounting that can be ignored.

The blackbody nature of the \cmb\ is realized as a requirement on the form of the $\mbA(ii)$'s.  If $\mbS(i)^\text{T}$ takes the form
  $(\mbSc(i)^{\text{T}},\mbF(i)^{\text{T}})$, with \mbSc\ for the 
  \cmb\ component(s) and \mbF\ for the foregrounds, then $\mbA(ii)$ must
  take the form
\ba
\mbA(ii)=\left( \begin{array}{c|*{2}{c}}
\mbe &?&?\\
\vdots&?&? \\
\end{array} \right),
\labeq{aform}
\ea
where \mbe\ is a ``tall'' matrix whose width equals the number of
components of $\mbSc(i)$ and whose elements are ones or zeroes (since temperature-type maps are assumed to be in thermodynamic
temperature units).
If a channel is polarization-sensitive, then its
corresponding part of \mbe\ will be a ``small'' identity matrix (with \mbi\
denoting identity matrices of various sizes from now on).  If a channel is
independent of the \cmb\, its corresponding part 
of \mbe\ will be a ``small'' (row) matrix of zeroes (all ``zero''
matrices will be denoted \mbzero).  Such a channel is still useful in the
analysis because it potentially responds to some of the foreground signals and 
thus carries information about them that can be used in extracting the \cmb.  
  The 
part of \mba\ marked ? encodes the spectral indices of the foreground
signals.

Because some physical foregrounds (e.g.\ galactic synchrotron
radiation) are known to 
have a spectral index that varies across the sky, one might expect
that our $\mbA(ii)$ should indeed be a function of the element $i$.
However, an alternative procedure is possible if we have many
frequency channels; we can take $\mbA(ii)=\mba$ for all $i$, and describe a
single physical foreground with 
varying spectral index as two or more correlated
effective signal components which have linearly independent corresponding
columns of 
$\mba$.    

From now on we will take \mba\ to be a square $n$-by-$n$ matrix, assuming there
to be as many signal components as we have frequency channels.  As we'll be marginalizing
out the foreground signals, it seems reasonable to allow for as
many of them as we might need to describe the data on top of the
noise. 

We are aiming towards using Bayes' Theorem to get a probability for a \cmb\
model (or perhaps a \cmb\ sky map) given the data, marginalized over
foregrounds.  Let us set down the relevant conditional probabilities
that we shall need.  Firstly,  
\ba
p(\mbS | \txtcm, \txtfm) 
= p( \mbSc | \txtcm) . p(\mbF | \txtfm)
\ea
by the independence assumption, where ``f.g.'' is short for foreground.  Here
$\mbSc$ is the \cmb\ signal (temperature and perhaps
polarization) and \mbF\ 
are the foreground signals, as introduced above.  
Typically the \txtcm\ is describable in terms of the \cmb\ power
spectra, but in
principle we could consider non-gaussian corrections too.  In the
gaussian case,
\ba
p(\mbSc | \txtcm) = \frac{1}{\sqrt{\left|2\pi \mbC\right|}} e^{-\mbSct \mbCm \mbSc /2}.
\ea
For
isotropic models in harmonic space, \mbC\
is diagonal in $l$
and conventionally expressed in terms of \cl's (i.e.\ $ \<a^P_{lm}
a^{Q}_{l'm'}\>=C_l^{PQ} \delta_{l,l'} \delta_{m,-m'}$ where $P$ and $Q$ denote
Stokes parameter type.).  

Next, the probability of the data
given the signals and the mixing matrix is
given by the probability of obtaining the noise realization
$\mbM=\mbX-\mbA\mbS$:
\ba
p(\mbX | \mbS, \mbA)=\frac{1}{\sqrt{\left| 2\pi \mathbf{N}\right|}} e^{-(\mbX-\mbA\mbS)^\text{T}
  \mbN^{-1} \(\mbX-\mbA\mbS\)/2}.
\ea 
Here we have used maximum entropy to assign a gaussian distribution
to the noise (see \cite{Jaynes}) and have decided not to add in and
marginalize over any 
non-gaussian corrections we could imagine (which would complicate the
analysis).  In some cases we might wish to consider ``inverse noise
matrices'' $\mbN^{-1}$ that are not actually the inverse of anything
due to having had say monopole and dipole components, point sources,
regions of strong galactic emission or instrumental effects
projected out of them.  (Such projections ensure that a model is not
penalized if the data and model ``disagree'' along such directions.)
In this case the determinant factor will not exist but seeing as it is
independent of the model it can be safely ignored.  Note that in many
applications where the sky map data is actually derived with a
maximum-likelihood approach from a
timestream it is in fact the inverse noise covariance matrix and not
the noise covariance matrix itself that naturally emerges.    
  
Now we need some priors.  For the \cl's, we might employ a
Jeffreys'-type prior $l$ by $l$, or alternatively assume they come from a
known cosmological model which itself has priors on its parameters.
In a middle-ground position we might incorporate a positive correlation
between neighbouring \cl's, expecting them to be given by integrals
over related transfer functions times the same 
underlying three-dimensional primordial power  
spectrum~\cite{Efstathiou:2003wr}.  
As simple test cases we could assume that the \cmb\ is gaussian
white noise on the sky with some unknown amplitude or that
the \cmb\ is scale-invariant with unknown amplitude. 
We shall see that the ``white noise'' assumption leads directly to the
\ilc\ approach as used by the \wmap\ team.  

We've already effectively imposed a prior that the mixing matrix \mbA\
is
block-diagonal in pixel space and have argued that it can be taken to be
isotropic across the sky if we have enough sky maps to work with.
(If we do relax the isotropy assumption on
\mba\ then we should parametrize a slow variation of \mba\ across the sky
with a suitable probability distribution over the parameters; see the
recent work of \cite{Kim:2008zh} for an application of this type of approach
to \ilc\ map making.)  One
might be concerned whether it is most natural to put a measure on
\mba\ or on its inverse, since the former is perhaps more natural when
thinking about the spectral response of physical foregrounds whereas
the latter is heuristically at least
what is needed to go from the data to the \cmb.  In fact we shall
develop a measure that is invariant under inversion.  A matrix measure
naturally involves some power of the determinant of the matrix, and
from $\mbam\mba=\mbi$ one can see that
$\delta\mbam=-\mbam\delta\mba\mbam$ and hence that $\partial (\mbam)
/\partial (\mba)=|\mbam|^{2n}$.\footnote{The latter is most easily seen by
considering the two linear transformations on $\delta\mba$ that
aggregate to give $\delta\mbam$, i.e.\ bracketing the latter as $-(\mbam
\delta \mbam ) \mbam$, using the chain
rule for Jacobians and performing row and column interchange
operations on each resulting $n^2$-by-$n^2$ matrix.} 
So a measure of $|\mba|^m d \mba$ transforms into a measure
$|\mbam|^{-m-2n} d\mbam$ for some power $m$.  Full form symmetry is
achieved if we choose $m=-m-2n$, or $m=-n$.  This choice yields
the Haar measure on $GL(n,R)$, the group of symmetries of a real
$n$-dimensional vector space, and so seems very natural.  This measure
is also scale-invariant as well as inversion-invariant.  From this
base we now consider our constraints on the mixing matrix from our taking the \cmb\ to be
black-body and to be independent from the foregrounds.  Start from the matrix equation
$\mbam \mba = \mbi$.  If we write \mbam\ as
\ba
\mbam=\left(
\begin{array}{*{1}{c}}
\mbwt \ldots\\
\hline
\mbut \ldots\\
\end{array} \right),
\labeq{ainv}
\ea
then we must have
\begin{subequations}
\labeq{amconstraints}
\ba
\mbwt \mbe&=&\mbi, \\
\mbut \mbe&=&\mbzero, 
\ea
\end{subequations}
where \mbw\ and \mbu\ are appropriate ``tall'' matrices.  From this we can see
that delta
functions required in our measure on \mba\ to force the first columns of $\mba$ to
equal \mbe\ correspond to the term $|\mbam|^{-1}
\delta(\mbwt\mbe-\mbi)\delta(\mbut\mbe)$ in a measure for $\mbam$.
With the form of \mbe\ as discussed above,
Eqs.~\refeq{amconstraints} render $|\mbam|$ independent of
$\mbw$.\footnote{This can be seen by using Eqs.~\refeq{amconstraints}
  to substitute for the variables appearing in the first \ncmb\
  columns of \mbam, then appropriately adding in the other columns to
  make the first ones equal $(\mbi,\mbzero)^\text{T}$ in the determinant
  calculation.}  Hence the components \mbw\
needed to extract the \cmb\ only actually appear in the delta function
$\delta(\mbwt\mbe-\mbi)$ and not in any prefactor in a base measure on
$\mbam$.\footnote{If we do have useful information on the foreground spectra
we would presumably first encode this in \mba\ and then consider
transforming to \mbam.  If this information is actually coming from
other non-\cmb\ sky maps, it might perhaps be easier to perform a combined
analysis including these maps as discussed above but with a larger \mba.}  

The final prior is for the foregrounds.  We shall take them to be
gaussian, with auto- and cross- $l$ by $l$ correlations.  The gaussian
assumption is not
ideal; indeed, the \wmap\
K- and KQ- sequence of masks are actually constructed by excluding
pixels whose temperatures fall into the
skewed high tail of the one-point temperature distribution function.
However, by taking the 
inverse covariance to 
zero, a gaussian model does at least provide an
analytically-controlled method of approaching a flat 
prior on the foregrounds.  A flat prior might be particularly appropriate for
polarization, given our limited physical understanding of galactic
polarization, and so is what a maximum entropy argument would yield.\footnote{As with for the mixing matrix, we could take any data sets that
do give us extra information into account by directly including them
in the analysis.}

\section{\labsec{withnoise}Probabilities for CMB Power Spectra with Noise}

We now move on to our main interest, the calculation of probabilities
for cosmological models, expressed in terms of their predicted \cmb\
power spectra, in presence of general (anisotropic and correlated)
instrument noise and foreground contaminants.

We shall take a limit towards a flat prior on the foregrounds, starting
from a gaussian
prior on them.  As we shall see, this limit conveniently decouples the
foreground-related parts of the inverse mixing matrix from the rest of
the problem.  We'll call the ``grand'' covariance matrix, of \cmb\ and
foregrounds combined, \mbG, and under our stated assumptions this fully
specifies our \cmb\ and foreground models.

We have:
\ba
p(\mbX|\mbam,\mbG) &=& \int d\mbS p(\mbX,\mbS|\mbam,\mbG) \\
&=& \int d\mbS p(\mbX|\mbS,\mbam,\mbG) p(\mbS| \mbam,\mbG) \\
&\propto& \int d\mbS 
\frac{
  e^{-(\mbX -\mbA \mbS)^T\mbNm(\mbX -\mbA \mbS)/2} }
  {\sqrt{|2\pi  \mbN|}} \sqrt{|\mbGm/2\pi|} e^{-\mbS^T \mbGm  \mbS /2} 
\ea
and to do the integral over \mbS\ it is useful to change variables to
$\mbU\equiv \mbA \mbS$, generating a $|\mbA|$ determinant factor.
With the mixing matrix being uniform across the sky, \mbA\ decomposes
into \mba's down the diagonal and so
$|\mbA|=|\mba|^\nelem$.  
Note that the right-hand-side has been written so as to not explicitly
depend on \mbG, only on \mbGm, to make the limiting process clearer.
We can perform the gaussian
integral over \mbS\ and use the result in Bayes' theorem, in conjunction with
our 
priors discussed in \refsec{priors}, to find:
\ba
p(\mbam, \mbG| \mbX) 
&\propto& p(\mbX | \mbam,\mbG) p(\mbam,\mbG) \\ 
&\propto& 
p(\mbG)
\frac{\sqrt{|\mbGm|}}{|\mbam|^{-\nelem}}
\frac{\delta(\mbwt\mbe-\mbi) \delta(\mbut \mbe)}{|\mbam|^{n+1}} 
\frac{e^{+\mbXt \mbNm (\mbNm+\mbAmt \mbGm \mbAm)^{-1} \mbNm
    \mbX/2-\mbXt \mbNm \mbX/2} }
{\sqrt{|\mbNm+\mbAmt \mbGm \mbAm|}}. 
\labeq{postforamandg}
\ea
To do this integral we needed $\mbNm+\mbAmt \mbGm \mbAm$ to be invertible.
We also see that as long as the number of elements is much larger than the
number of frequency channels, the 
$|\mbam|^{n+1}$ term from our prior on
\mbam\ becomes insignificant and could be ignored relative to the
$|\mbam|^{\nelem}$ term from the likelihood.  We keep both however, defining $N\equiv \nelem-n-1$.   

It is convenient to take the flat-foreground-prior limit $\mbFm \rar
0$ at this
point.  First note that $|\mbGm|=|\mbCm||\mbFm|$ since \cmb\ and
foregrounds are assumed independent.   By taking the limit in
the same manner for the different \cmb\ models that we are
considering, we can safely ignore the vanishing $|\mbFm|$ factor
(alternatively we can say that we are only interested in likelihood
ratios between models, in which case the factor vanishes).  

Next, we consider $\mbAmt \mbGm \mbAm$.  Written out more explicitly,
this takes the form
\ba
\(
\begin{array}{ccc}
\mbamt \mbGm (1,1) \mbam & \mbamt \mbGm (1,2) \mbam & \\
\mbamt \mbGm (2,1) \mbam & \mbamt \mbGm (2,2) \mbam & \\
&&\ddots
\end{array}
\)
\ea
where we have decomposed $\mbGm$ into element-labelled blocks,
$\mbGm (ij)$ corresponding to the block of \mbGm\
relating to elements $i$ and $j$.  As $\mbFm \rar 0$, $\mbamt \mbGm
(ij) \mbam \rar \mbw \mbCm (ij) \mbwt$.  So, introducing the 
$n\nelem$-by-$\ncmb \nelem$ matrix $\mbQ$:
\ba
\mbQ \equiv
\(
\begin{array}{cccc}
\mbw &\mbzero &\mbzero& \cdots \\
\mbzero &\mbw & \mbzero& \\
\vdots &\vdots &\vdots&
\end{array}
\),
\ea
we have
\ba
\mbAmt \mbGm \mbAm \rar \mbQ \mbCm \mbQt.
\ea
Thus the probability factors into a term depending on $\mbw$ and $\mbC$
alone and a term depending on $\mbu$ alone (the $\delta (\mbut
\mbe)/|\mbam|^{-N}$ part).   When we integrate
over $\mba$, or equivalently over $\mbw$ and $\mbu$, to obtain the
marginalized probability for the \cmb\ model, the $\mbu$ integral thus
factors off to give a multiplicative constant that can be ignored.  

Hence we obtain an effective probability for a \cmb\ model
of
\ba
p(\mbC | \mbX) \propto \int d \mbw \delta (\mbwt \mbe -\mbi)
\frac{e^{+\mbXt \mbNm (\mbNm+\mbQ \mbCm \mbQt)^{-1} \mbNm \mbX/2 
-\mbXt \mbNm \mbX/2} }
{\sqrt{|\mbC| |\mbNm+\mbQ \mbCm \mbQt|}}.
\labeq{effcmbprob}
\ea
The integral 
over \mbw\ will be performed using the saddle-point method developed in~\refapp{saddle}.  

Before integrating though, there may exist opportunities to
significantly simplify \refeq{effcmbprob}, depending on the
invertibility properties of the matrices remaining.  If so, the size
of subsequent matrix operations is substantially reduced, 
easing the computational burden.  For the rest of
this section we consider the simplest case 
of both $\mbNm$ and $\mbCm$ being invertible; more complicated cases
are deferred to subsections below.  Following the discussion in
\refapp{woodbury}, we are able to apply formulae~\refeq{wood}
and \refeq{wooddet} to simplify \refeq{effcmbprob} to
\ba
p(\mbC | \mbX) \propto p(\mbC) \int d \mbw \delta (\mbwt \mbe -\mbi)
\frac{e^{-\mbXt \mbQ (\mbC+\mbQt \mbN \mbQ)^{-1} \mbQt \mbX/2}}
{\sqrt{|\mbC+\mbQt \mbN \mbQ |}}.
\labeq{simplecmbprob}
\ea
Note how the size of the matrix that needs to be inverted and have its
determinant found has reduced from $n\nelem$-by-$n\nelem$
to $\ncmb\nelem$-by-$\ncmb\nelem$, where \ncmb\ is the number of
\cmb\ Stokes 
parameters under consideration, saving $O((n/\ncmb)^3)$ in required computation.

Comparing with Eq.~\refeq{theint}, we have
\ba
S= \frac{1}{2} \( \ln |\mbC+\mbQt
\mbN \mbQ| + \mbXt
    \mbQ (\mbC+\mbQt \mbN \mbQ)^{-1} \mbQt \mbX \),
\labeq{newexp}
\ea
and by varying with respect to $\mbw$ the appropriate first and second
derivatives for $S$ can be read off for use in \refeq{update}:
\ba
\delta_\mbw S &=& \tr \delta \mbwt \left\{ \mbN (ij) \mbw
\left( \mbDm 
(ji)
- \mbDm (jm) \mbX(m) \mbXt (n) \mbDm (ni) \right) 
+ \mbX(i) \mbXt (j) \mbw \mbDm (ji) \right\} \\
\delta^2_\mbw S &=&  
\tr \delta \mbwt \mbN(ij) \delta \mbw \left( \mbDm(ji) 
- \mbDm (jm) \mbX(m) \mbXt (n) \mbDm (ni) \right)
/2 \nonumber \\ 
&+&\tr \dmbwt \mbN(ij) \mbw \mbDm (jk) \dmbwt \mbN(ki) \mbw \nonumber \\
&+& \tr \dmbwt \mbN(ij) \mbw \mbDm (jk) \mbwt \mbN(ki) \dmbw \nonumber\\ 
&+& \tr \dmbwt \mbN(ij) \mbX(i)\mbXt(j)\mbw \mbDm(ji) /2 \nonumber\\
&-& \tr \dmbwt \mbX(i) \mbXt(j) \mbw \mbDm(jk) \left(\dmbwt \mbN (km)
\mbw +\mbwt \mbN (mn) \dmbw \right) \mbDm(ni)\nonumber \\
&+&\tr \dmbwt \mbN (ij) \mbw \mbDm(jk) \dmbwt \mbN(km)\mbw
 \mbDm(mn) \mbwt \mbX(n)\mbXt(p)\mbw \mbDm(pi) \nonumber\\
&+&\tr \dmbwt \mbN(ij) \mbw \mbDm(jk) \mbwt \mbN(km) \dmbw \mbDm(mn)
\mbwt \mbX(n) \mbXt(p) \mbw \mbDm(pi) /2 \nonumber\\
&+&\tr \dmbwt \mbDm(ij) \dmbw \mbN(jk) \mbw \mbDm(km) \mbwt \mbX(m)
\mbXt(n) \mbw \mbDm(np)\mbwt\mbN(pi) /2
\ea
where we have defined $\mbD$ to equal $\mbC+\mbQt \mbN \mbQ$ and
assumed implied summation over repeated element indices.  (To
obtain full equivalence with Eq.~\refeq{update} we need to map from
the matrix \mbw\ here to the vector $x$ there. This can be done
explicitly by ``vectorization'' of \mbw\ if required, as discussed in
\refapp{veckron}.) 
The numerical calculation time is
dominated by the required $O((\ncmb\nelem)^3)$ inversion of \mbD.   So as
long as the initial guess for \mbw\ is reasonable\footnote{A suitable starting
  place should be obtainable from the ``noise-free'' solution evaluated for a
  reasonable fiducial model, as derived in
  \refsec{nonoise}.}, one should be able
to converge to the saddle-point expression for \mbw\ in a few
$O((\ncmb \nelem)^3)$ steps.  Having converged to a saddle point solution
$\mbw\subsp$,  we evaluate the exponent \refeq{newexp} there,
calculate a prefactor\footnote{This prefactor coming from the integral over \mbw\ is helping to encode the
uncertainties from the foreground separation into the likelihood; this term
would be missed in any approach that attempts to use a standard \cmb\
likelihood formula applied to some best-fit linearly combined map and
effective noise matrix.} involving second derivatives of $S$ at the
saddle point as discussed in \refapp{saddle} and multiply by
any prior in order to finally 
obtain the posterior probability for the \cmb\ model in question.

\subsection{Noise Matrices with Projections}

Often one wishes to project additional degrees of freedom out of a
formally invertible inverse noise matrix in order to render our
posterior probabilities insensitive to certain complications with the data
that would otherwise be unaccounted for.  For example, one might
project out monopole and dipole contributions of the maps from an
experiment like \wmap, the monopole because the experiment is
basically differential and the dipole because of the earth's motion
relative to the \cmb.  Or one might project
out regions
of strong emission from the galaxy because one does not expect the
relatively simple foreground model used to be accurate there.  The
\wmap\ team also project out a ``transmission loss imbalance'' mode
from their maps.  In some
cases one can sometimes 
just reduce the dimensions of the matrices involved; e.g.\ one might
just use the $l>1$ modes in harmonic space to forget about the monopole
and dipole, or for a
galactic cut working in pixel space one might simply only consider
data from pixels outside of the cut.  However, when say both a
monopole and dipole projection and a galactic cut are made, no such
simple approach is possible, and one must proceed as follows.

Each of the \nproj\ modes to be projected out are expressed as a
column vector, and
these column vectors are arranged into an $n\nelem$-by-\nproj\ matrix \mbP.
Then, $\mbNm$ is replaced by:
\ba
\mbNm \rar\ \overline{\mbNm} |_{\mbP} \equiv \mbNm-\mbNm \mbP \(\mbPt \mbNm
\mbP\)^{-1} \mbPt \mbNm.
\ea
(Note that the normalization of the modes cancels out in the formation
of $\overline{\mbNm}|_{\mbP}$.)  $\overline{\mbNm}|_{\mbP} $ has
been constructed so that when it 
is multiplied into any linear combination of the modes to be projected
out it returns zero.  Hence any discrepancy between the data and a
potential signal along these modes is not penalized in the likelihood.       

Eq.\ \refeq{effcmbprob} becomes
\ba
p(\mbC | \mbX) \propto \int d \mbw \delta (\mbwt \mbe -\mbi) 
\frac{e^{+\mbXt \overline{\mbNm}|_{\mbP} (\overline{\mbNm}|_{\mbP}+\mbQ \mbCm \mbQt)^{-1}
    \overline{\mbNm}|_{\mbP} \mbX/2  
-\mbXt \overline{\mbNm}|_{\mbP} \mbX/2} }
{\sqrt{|\mbC| | (\overline{\mbNm}|_{\mbP}+\mbQ \mbCm \mbQt)|}},
\labeq{projeffcmbprob}
\ea
which, after some work involving \refeq{wood} and \refeq{wooddet} and
keeping \mbN\ to refer to the assumed-invertible formal noise matrix,
reduces to:
\ba
p(\mbC | \mbX) \propto p(\mbC) \int d \mbw \delta (\mbwt \mbe -\mbi)
\frac{e^{-\mbXt \mbQ \left. \overline{\left(\mbC+\mbQt \mbN
      \mbQ\right)^{-1}}\right|_{\mbQt\mbP} 
    \mbQt \mbX/2}} 
{\sqrt{|\mbC+\mbQt \mbN \mbQ ||\mbPt\mbQ (\mbC+\mbQt \mbN \mbQ)^{-1}
    \mbQt \mbP| }}.
\labeq{projcmbprob}
\ea
We can now expand the delta function as a gaussian and integrate as
above.  Additional terms in the first and second derivates of the
exponent come from the projection, further complicating the
formulae, but the numerical time of the calculation is still dominated
by the inversion of $ \mbC+\mbQt \mbN \mbQ$.  

There is some subtlety about which projections can be handled in this
manner.  For example, with temperature maps alone, one cannot
independently project out all 
measurements associated with a given element; the
second determinant under the square root turns out to be proportional
to $|\mbw \mbwt|^{\nelem}$ and is thus singular.  (This is happening perhaps
because as the foregrounds are being marginalized out and the dimensions effectively reducing from $n 
\nelem$ to $ \nelem$ the previously orthogonal projections might
be 
becoming linearly dependent.)   
Potentially including polariation again, one can however project out
any response to 
the \cmb\ in 
the $i$'th element say with $\mbP(i)$ equal to $\mbe$ and all other
entries in \mbP\ zero.   With the help of the delta
function $\mbQt \mbP (i)$ becomes equal to the identity
matrix \mbi.  Finally, one obtains just
the result one would have gotten by forgetting about the element in
question in the first place and starting with a reduced problem with
only $\nelem-1$ elements rather than \nelem\ elements, as can be seen
using rules for the determinants of blocked matrices and
their inverses.\footnote{By 
  only using certain columns of $\mbe$ in $\mbP(i)$ one can in fact
  project out individual Stokes components 
  of the \cmb, paving the way for independent temperature and
  polarization masks.}  By combining many such projections into a wide
projection matrix one can
form a mask, or by considering an appropriately weighted
vector sum of such projections one can project out spatially coherent
modes such 
as the monopole or dipole from pixellized data.
  
\subsection{The general case}  

Finally we mention the case when the inverse noise matrix is
completely general.  Now, we cannot use the Woodbury formula to
simplify the exponential in \refeq{effcmbprob}. Rather, we have to use
the effective exponent from \refeq{effcmbprob} as is and directly proceed
with the replacement of the delta function with a gaussian.  It is in fact
very appealing to work like this directly with inverse noise matrices and
inverse-noise-weighted maps, as these are what naturally emerge out of a
maximum-likelihood analysis of timestream data.     However, the practical cost
is that one now has to invert full $n \nelem$-by-$n \nelem$
matrices for each \cmb\ model considered. 

\section{\labsec{nonoise}The no-noise limit}

It is interesting to consider what happens if the noise is negligible.
Starting from \refeq{simplecmbprob}, we can take the limit $\mbN \rar0$ to
obtain:
\ba
p(\mbC | \mbX)_\text{no noise} \propto \int d\mbw \delta (\mbwt\mbe-\mbi) 
\frac{p(\mbC)}{\sqrt{\left| \mbC \right|}} e^{-S'},
\ea
with exponent $S'= \vcwt  \hatR \, \vcw /2$ (with ``vec'' indicating
vectorization as discussed in \refapp{veckron}, and using
Eq. \refeq{veckronid}) 
where we have defined 
\ba
\hatR(\mbC) = \sum_{ij} \mbCm(ij) \otimes \mbX(i) \mbXt(j).
\ea
Here $\otimes$ denotes the Kronecker product, discussed in
\refapp{veckron}.  Notice that $S'$ is now quadratic in the components of \mbw\ and
hence the saddle point integration over \mbw\ is exact.
Re-expressing the $\mbwt \mbe=\mbi$
constraint as 
\ba
(\mbi \otimes \mbet) \vcw=\vci,
\ea
Eq.~\refeq{quadraticsoln} 
gives the saddle point for \vcw\ at:
\ba
\vcw\subsp =\hatRm (\mbi \otimes \mbe) \((\mbi \otimes \mbet)  \hatRm
(\mbi \otimes \mbe) \)^{-1} \vci 
\labeq{nonoisew}
\ea
(``s.p.'' for saddle point). Using \refeq{quadraticaction}, the posterior for the \cmb\ model is
\ba
p(\mbC | \mbX) \propto \frac{p(\mbC)}{\sqrt{\left| \mbC \right|}}
\frac{1}{\sqrt{ \left|\hatR\right| \left|(\mbi \otimes \mbet)  \hatRm
(\mbi \otimes \mbe) \right|}} e^{- \vcit
  \((\mbi \otimes \mbet)  \hatRm 
(\mbi \otimes \mbe) \)^{-1} \vci /2}.
\labeq{nonoiselike}
\ea

Eq~\refeq{nonoisew} bears a marked resemblance to the formula for the
weights in the \ilc\ approach and this will be expanded upon in the following
section.

\section{\labsec{ilc}Deriving sky maps and relations with the ILC procedure}

So far in this paper we have concentrated on deriving likelihood
functions for \cmb\ models rather than on producing \cmb\ sky maps. 
However, our procedure can of course be used to generate sky maps and
associated noise covariances including contributions from foreground
removal.    

Now $p(\mbS|\mbX)=\int d\mbA \,d\mbG\, p(\mbS,\mbA,\mbG|\mbX) \propto \int
d\mbA \,d\mbG\, p(\mbX|\mbS,\mbA,\mbG) \, p(\mbS,\mbA,\mbG)$ using Bayes'
Theorem.  Using the same conditional probabilities and priors as
above, again introducing $\mbU = \mbA \mbS$, and using Eq.~\refeq{spmean}, we obtain:
\ba
\< \mbSc(i) \> &\approx& 
\int d\mbC p(\mbC|\mbX) \mbwt\subsp(\mbC) \mbX(i) \labeq{generalc} 
\ea
along with a somewhat messier expression for $\< \mbSc(i) \mbSct(j) \> $
using Eq.~\refeq{spcov}.  In the no-noise limit, these
expressions become exact and $\mbwt\subsp(\mbC)$ and $p(\mbC|\mbX)$
are obtainable from
Eqs.~\refeq{nonoisew} and \refeq{nonoiselike} respectively.

To establish a direct link with \ilc\ let us try to construct the
\cmb\ temperature (and possibly 
polarization) maps, $\mbSc$.
Now, imagine we are working in pixel space in the no-noise limit and
we have a prior on the 
\cmb\ that it is sky-uncorrelated, i.e.\ it is gaussian white noise.
Then $\mbC(i,j)=\mbc \, \delta_{ij}$ say and \hatR\ reduces to $\mbcm
\otimes \hatx$, defining $\hatx \equiv \sum_i
\mbX(i) \mbXt(i) $.  Then $\hatRm$ is just
$\mbc \otimes \hatxm$ using \refeq{kroninv}. Eq.~\refeq{nonoisew}
simplifies to become 
independent of \mbc, and so, no matter what our prior on \mbc\ actually
is, we have
\ba
\<\mbSc(i)\>= \mbwt\subsp \mbX(i)
\labeq{ilcmap}
\ea
with 
\ba
\mbw\subsp = \hatxm \mbe \( \mbet \hatxm \mbe \)^{-1}.
\labeq{ilcw}
\ea
For temperature alone, in which case \mbe\ is just
$(1,\ldots,1)^\text{T}$, this is exactly the \ilc\ result. So
Eqs.~\refeq{ilcmap} and \refeq{ilcw} are the natural generalization of
\ilc\ when treating 
temperature and polarization in a unified manner.    
Typically the \ilc\ procedure is described
as 
choosing the linear combination of channels that has minimum variance
whilst still retaining unit response to the 
\cmb.  Our Bayesian derivation here (albeit with flat priors on the
foregrounds and a very strong, and obviously incorrect, prior on the
\cmb) here gives an alternative, more 
insightful, perspective.  For example, we see that one might not need
to correct  
for ``cosmic covariance'' \cite{Hinshaw:2006ia,Chiang:2007rp}; the \ilc\
coefficients already give the mean maps.  We might replace the ``white
noise \cmb'' prior with
one based on a fiducial model up to an overall
amplitude and thus derive ``improved'' \ilc\ coefficients that
correctly take into account spatial correlations in the \cmb.  A very simple
case appropriate for low-$l$ would be to take the \cmb\ to be
scale-invariant, $C_l \propto 1/(l(l+1))$, and work in harmonic space,
calculating \hatR\ with the appropriate weights.  

As mentioned below Eq.~\refeq{generalc}, we can calculate a measure of
the foreground-induced pixel-pixel covariance of our maps.  For
clarity we shall do this for temperature alone in the no-noise
limit.  Then $\mbSc(i)$ is simply $t(i)$, the temperature in the $i$'th
pixel.  With the ``white noise'' \cmb\ prior, \mbc\ is given by the
single number \dtsq.  With a  
Jeffreys' prior 
on \dtsq\, one finds
\ba
\<\<t(i)t(j)\>\>=\frac{1}{\nelem-n+1} (\mbX(i)-\<t(i)\>
\mbe)^\text{T} \frac{\hatxm}{\mbet \hatxm \mbe} (\mbX(j)-\<t(j)\>
\mbe).   
\ea
This is a very reasonable result; uncertainties are given by a
quadratic form on differences away from a blackbody, with the
quadratic form determined by the covariance of the sample.

One can attempt generalizing the \ilc\ procedure in other ways.  For
example, staying with temperature alone, one could keep the ``white
noise'' \cmb\ prior, but try to 
add in instrument noise at some level, taking it say to be
sky-uncorrelated and isotropic, described by the same matrix \mbn\ at
each pixel.  Then the
joint likelihood for \mbw\ and \dtsq, with a flat prior on \dtsq,
turns out to be:
\ba
p(\dtsq,\mbw|\mbX) \propto \frac{\delta(\mbwt
  \mbe-1)}{\sqrt{\dtsq+\mbwt \mbn \mbw}^N} 
e^{-\frac{1}{2}\frac{\mbwt \hatx \mbw}{\dtsq+\mbwt \mbn \mbw}},
\labeq{jointnoise}
\ea
and the joint maximum likelihood point for \dtsq\ and
\mbw\ together occurs at the just same value for \mbw, namely $\hatxm
\mbe /(\mbet \hatxm \mbe)$, as in the noise-free case, with unchanged
variance for \mbw.  This is at odds with the idea of subtracting the noise
contribution \mbn\
from the total variance \hatx\ (which might seem plausible on the grounds
of focussing on the signal) in 
deriving the \ilc\ coefficients.  

\section{\labsec{extensions}Possible Extensions, Practical Considerations and Conclusion}

While the scheme presented above treats foregrounds very generally,
there is scope for further extension.  One possibility is to allow for
a slow variation of the mixing matrix across the sky.  In fact it
would be 
technically easiest to perform a low-order spherical harmonic
expansion of the inverse pixel-space mixing matrix; then $\mbwt (\mbx)
\mbe =\mbi$ would reduce to $\mbwt_{00} \mbe=\mbi$ and $\mbwt_{lm}
\mbe=\mbzero$ for $l>0$.  Ref.~\cite{Kim:2008zh} applies an analagous
approach to map making, allowing the \ilc\ coefficients to vary across
the sky.  Notice that this is distinct from the harmonic approach of
\cite{ 1996MNRAS.281.1297T, Tegmark:2003ve} and is rather more
physical when looked at from a component separation approach; the $l$-by-$l$
analysis there effectively 
imagines the data from given direction to come from a convolution of
signals around that direction, whereas \cite{Kim:2008zh} and the suggestion
here consider the data in a given direction to come from signals in
that direction alone.   

Another natural variation would be to
relax the flat prior on the foregrounds and rather try to marginalize
over power spectra for the foregrounds.  Now the separation achieved
above between \cmb\ and foregrounds would not occur and one would have
to think about possible priors on the foreground part of the inverse 
mixing matrix (luckily we have seen that the base prior on \mba\ rapidly
becomes irrelevant as the number of elements increases).  The underlying model
would be close to     
that used in the Gibbs sampling approach of~\cite{Eriksen:2007mx}, and
in the context of map making one would be left with a
scheme very similar to that of~\cite{Delabrouille:2002kz} (but with
the advantages of working with \mbam\ rather than \mba).  Now one of the
distinguishing features of the foreground signals is that they are
non-gaussian, and one could hope to exploit this if one could develop
a plausible non-gaussian correlated probability distribution for the
foregrounds.   As mentioned earlier, one might also consider
non-gaussian corrections to the \cmb\ itself, but in this case at
least we suspect them to be small.  

A conceptually simple change would be to relax the assumption that the
mixing matrix \mba\ is square, if say one is confident in the number
of physical foreground emission processes operating.  However, many of the
manipulations of our approach depend technically on the invertibility of \mba,
so this change would be 
difficult to implement and would only be trustworthy if any neglected
foreground components (e.g.\ coming from non-modelled spectral index
variations) were below the level of the detector noise. 

Let us now move on to discuss certain practical considerations in implementing
a scheme such as that described here.  An obvious difficulty with a
straightforward implementation is the usual one confronting \cmb\
analysis, namely the need to numerically invert large matrices
corresponding to 
the large size of the data sets involved.  Although the cosmic signal is naturally band
limited, the cutoff is too high for a full resolution
analysis over a good fraction of the sky to be practical.\footnote{Of
  course one might  
  be able to apply the scheme directly to data from high resolution
  ground based 
  experiments that only focus on small patches of the sky.}  
With simplifying assumptions however, some progress may be
possible.\footnote{For example, one might work in harmonic space and
  assume that the detector noise, along with the signal, is diagonal in this
  basis.  Then many of the matrix operations simplify.  Still
  implementing a general cut using the projection technique, the
  computational burden might be reduced by the ratio of the cube of the
  number of modes 
  projected out to the cube of the number of elements.  A limited
  number of further
  corrections to the noise matrix might also be efficiently
  implementable using Woodbury formula techniques.  Alternatively one
  might consider perturbative corrections to the noise model away from the
  (masked) isotropic case.}  
  In any case, a full resolution analysis is probably not even necessary.
  Instead, one might perform
a split analysis, 
treating large and small angular scales in a different manner (see
\cite{Efstathiou:2003dj} for a discussion in the absence of
foregrounds).  While the large angular scales (where some of the most
interesting new results might manifest themselves and where particularly
B-mode polarization might have some signal relative to detector noise
for Planck) might
be dealt with accurately by an application of the method of this paper, an
heuristic approximation to the method (or a different scheme altogether) might
be all that is needed
for high $l$.  Of course the problem will not precisely factorize into a large
scale part and a small scale part and so the coupling will have to be
taken into 
account or shown to be negligible.   In addition, there are issues with the
production and 
interpretation of low resolution maps and their (inverse) noise
matrices (see the description in~\cite{Jarosik:2006ib} of the procedure used
by the \wmap\ team), with difficulties due to aliasing and the finite size of
the pixels themselves.  In this work we have assumed that the signal has been
deconvolved from the 
beam; the appropriate noise matrices must be obtained by transforming those
appropriate to the common case of solving only for the beam-smoothed
signal.   Any beam uncertainties should be incorporated in the noise
matrices also.

Often the time-ordered data is discretized into a
map on the sphere using some pixelization scheme as opposed to going directly
to harmonic space.   Even if all
signals are strictly band-limited, with non-zero spherical
harmonic coefficients only for $l\leq \lmax$ say, fully describing such
signals in pixel space requires more than $(\lmax+1)^2$ pixels.  One then has
to be careful about the validity of certain matrix operations that need to be
performed.  For example,
the inverse transformation from an unconstrained map in pixel space to
harmonic space is not well-defined.  Similarly, the pixel-space
signal covariance matrix is non-invertible.  The harmonic-space
inverse 
noise matrix can be obtained by transforming the pixel-space inverse noise
matrix though.   An important issue arises if a galaxy cut projection is to be
made: the mask itself needs be band-limited in order for pixel-space and harmonic-space
approaches to be equivalent.  

While all of these complications need to be investigated and accounted for if
necessary, it is not likely that they will lead to a fundamental problem in
the scheme defined here for cosmic information extraction in the presence of
foregrounds, just as they have not prevented standard cosmological analysis of
\cmb\ data.      

In conclusion, this paper has described a prescription for the analysis of
multiple sky 
maps for cosmological information in the presence of both detector noise and
foregrounds.  The prescription takes into account uncertainties in the
foregrounds by modelling the foregrounds in a very general way and then
marginalizing over the foregrounds.  The noise can be very general, potentially
correlated over the sky and even correlated between frequency channels, and
include projections.  Work
is underway both   
to subject the prescription to extensive Monte Carlo testing and to apply it
to the \wmap\ data.  Assuming its performance is good the scheme should be
ideal for the analysis of the temperature and polarization
data from the 
forthcoming Planck satellite.

\acknowledgments

I thank George Efstathiou for directing my
attention to the problem of the foreground contamination of the \cmb, and
thank him, Mark Ashdown, Anthony Challinor, Antony Lewis and Francesco Paci
for helpful comments and discussions.  I am supported by STFC.

\appendix

\section{\labsec{saddle}Constrained Saddle Point Integrals}

This appendix explains the technique we use to evaluate constrained
integrals in a saddle-point approximation.  

Imagine we are performing an $n$-dimensional integral over variables
$x^i$ of some function $e^{-S(x)}$, with $m<n$
linearly-independent linear constraints of the form $c^{~k}_i x^i=d^k$
(summed over $i$) on the $x$'s applied, i.e.\
\ba
I=\int d^n x \delta(c^{~1}_i x^i-d^1) \ldots \delta(c^{~m}_i x^i-d^m)
e^{-S(x)}. 
\labeq{theint} 
\ea

First, we approximate the delta functions by a narrow gaussian,
described with some covariance matrix $D$ which will
eventually be taken to zero:
\ba
\delta(c^\text{T} x -d ) \equiv \delta(c^{~1}_i x^i-d^1) \ldots \delta(c^{~m}_i
x^i-d^m) 
\sim
\frac{1}{\sqrt{\left| 2\pi D \right|}} e^{-(c^{~m}_i x^i-d^m)
  {D^{-1}}_{mn} (c^{~n}_i x^i-d^n)/2}.
\labeq{deltaapp}
\ea
Next, we find the saddle point of the entire integrand and approximate
the integrand as a gaussian around the saddle point.  The saddle point
is where the first derivative
\ba
\left. c D^{-1} (c^\text{T} x-d) \right|_i +S_{,i}
\labeq{firstderiv}
\ea
of minus the exponent is zero, 
and the matrix $M_\text{total}$ of second derivatives of minus the
exponent has components 
\ba
\left. c D^{-1} c^\text{T} \right|_{ij}+S_{,ij}.
\labeq{secondderiv}
\ea
We need to evaluate both exponent and prefactor terms to obtain the
full saddle-point approximation to \refeq{theint}.   At the saddle
point, it turns out that the ``delta-function'' part of the exponent
tends to zero as $D\rar 0$, as can be seen as follows.  Multiplying 
\refeq{firstderiv} by $c$ on the left at the saddle point (``s.p.'' in
formulae below) and
rearranging yields
\ba
c^\text{T} x-d=- D (c^\text{T} c)^{-1} \left. S_{,i}\right|_\text{s.p.} 
\ea 
and so the ``delta-function'' part is
\ba
\left. S^\text{T}_{,i}\right|_\text{s.p.}
(c^\text{T} c)^{-1}
D (c^\text{T} c)^{-1}\left. S_{,i}\right|_\text{s.p.}/2.
\ea
If there exists a sensible solution as $D\rar 0$, then $\left.
S_{,i} \right|_\text{s.p.}$
will only have a weak dependence on $D$ and so in the limit the above
term will vanish.

The prefactor comes from the gaussian integral over deviations away
from the saddle point, which is proportional to the reciprocal square
root of $M_\text{total}$ evaluated at the saddle point.   Splitting
$M_\text{total}$ into the piece $c D^{-1} c^\text{T}$ from the delta
function and 
the piece $M$ with components $\left. S_{,ij}\right|_\text{s.p.}$, we
have
$|M_\text{total}|=|M| |D^{-1}| |D+c~\text{T} M^{-1}
  c|$.  Combining with the $\sqrt{|D|}$ term in the
denominator of \refeq{deltaapp} and taking the limit, we obtain an
overall prefactor proportional to $1/\sqrt{|M| |c M^{-1} c^\text{T}|}$.   

Hence we obtain: 
\ba
I\approx \frac{1}{\sqrt{\left| M / (2\pi) \right| \left| 2\pi c^\text{T} M^{-1}
    c \right| }} e^{-S},
\labeq{spanswer}
\ea 
with all quantities evaluated at the saddle point.  Note that the
result depends only on the 
original integrand and its second derivatives but evaluated at the
saddle point of the combined integrand.
Additionally,
\ba
\int d^n x \delta (c^\text{T}x -d) \, x  e^{-S(x)}
&\approx& x\subsp I, \labeq{spmean}  \\
\int d^n x \delta (c^\text{T}x -d) \, x x^\text{T} e^{-S(x)}
&\approx&\(M^{-1}-M^{-1} c
\left(c^\text{T} M^{-1} c \right)^{-1} c^\text{T} M^{-1} 
+x\subsp x^T\subsp \right) I.
\labeq{spcov}
\ea
   
We still have to actually find the saddle point and typically this can be done
numerically with a Newton-Raphson-type approach as follows.  With the first
and second derivatives from formulae \refeq{firstderiv} 
and \refeq{secondderiv} above, 
the Newton-Raphson update is of the form:
\ba
\delta x =-M_\text{total}^{-1} \(c D^{-1}(c^\text{T} x-d) +\nabla S(x)\)
.\ea
With the saddle point of the limit being the limit of the saddle point
we can simply start from a point on the constraint surface and take the limit
$D\rar 0$ straight away to obtain the projected update formula:
\ba
\delta_\text{proj} x = -\left( M^{-1}-M^{-1} c
\left(c^\text{T} M^{-1} c \right)^{-1} c^\text{T} M^{-1}\right) \nabla S(x)
\labeq{update}
\ea
(``proj'' for projected).   
Iterating this will lead us to the saddle point, which can then be
used in \refeq{spanswer} to obtain the approximate value for the integral.

In the special case that $S$ is quadratic in $x$, the saddle
point can be solved for analytically and is at $x = M^{-1} c \(
D+c^\text{T} M^{-1} c \)^{-1} d$. So taking $D \rar 0$, we have
\ba
x_{\text{s.p.}}=   M^{-1} c \( c^\text{T} M^{-1} c \)^{-1} d
\hspace{.5cm} \text{(for $S$
  quadratic in $x$)}
\labeq{quadraticsoln}
\ea 
and hence
\ba
S\subsp= d^\text{T} \( c^\text{T} M^{-1} c \)^{-1} d /2\hspace{.5cm}
  \text{(for $S$ quadratic in $x$)}.
\labeq{quadraticaction} 
\ea
Furthermore, Eqs.\ \refeq{spanswer}, \refeq{spmean} and \refeq{spcov}
  become exact. 

\section{\labsec{veckron}Vectorization and Kronecker Products}

We here summarize vectorization of matrices and the
Kronecker product of two matrices.  See Appendix A of
\cite{Hamimeche:2008ai} for further identities and a recent discussion
of vectorization in the context of \cmb\ analysis.

The vectorization $\text{vec} (A)$ of an $m$-by-$n$ matrix $A$ is the
$mn$-by-1 column vector formed by stacking the columns of $A$ on top
of each other.  Explicitly, $\text{vec} (A)^\text{T}=
(A_{11},\ldots,A_{m1},A_{12},\ldots,A_{m2},\ldots,A_{1n},\ldots,A_{mn})$. 

The Kronecker product of an $m$-by-$n$ matrix $A$ and a $p$-by-$q$
matrix $B$ is an $mp$-by-$nq$ matrix, denoted $A \otimes B$, formed by
appropriately stacking copies of the $B$ matrix that have been
multiplied by the elements of A:
\ba
A \otimes B =
\begin{pmatrix}
A_{11} B & A_{12} B & \cdots & A_{1n} B \\

A_{21} B & A_{22} B &\cdots & A_{2n} B \\
\vdots &\vdots && \vdots \\
A_{m1} B & A_{m2} &\cdots &  A_{mb} B \\
\end{pmatrix}.
\ea
When the matrices are of compatible sizes such that the relevant
products exist, 
\ba
(A\otimes B)(C\otimes D)=(AC) \otimes (BD).
\ea
Hence,
\ba
(A \otimes B)^{-1} = A^{-1} \otimes B^{-1}.
\labeq{kroninv}
\ea
Also,
\ba
(A\otimes B)^{\text{T}}=A^{\text{T}} \otimes B^{\text{T}}.
\ea

A useful identity involving both vectorization and the Kronecker
product is
\ba
\tr (A^\text{T} BCD) = 
\text{vec}(A)^\text{T} (D^\text{T} \otimes B) \text{vec} (C).
\labeq{veckronid}
\ea

\section{\labsec{woodbury}Summary of Matrix Identities Used}

This paper makes extensive use of Sherman-Morrison/Woodbury-type
formulae (see e.g.\ \cite{numrec}) for matrix inverses and
their determinants.  Here we briefly show how these results may be
derived in order to understand how and when they may be applied.  We
start by considering two related decompositions of the same blocked matrix:
\ba
\begin{pmatrix} a & -u \\ v^\text{T} & c^{-1} \end{pmatrix} 
&=&
\begin{pmatrix} 1 & -u c \\ 0 & 1 \end{pmatrix}
\begin{pmatrix} a+u c v^\text{T} & 0 \\ 0 & c^{-1} \end{pmatrix}
\begin{pmatrix} 1 & 0 \\ c v^\text{T} & 1 \end{pmatrix} \\
&=&
\begin{pmatrix} 1 & 0 \\ v^\text{T} a^{-1} & 1 \end{pmatrix}
\begin{pmatrix} a & 0 \\ 0 & c^{-1} + v^\text{T} a^{-1}u \end{pmatrix}
\begin{pmatrix} 1 & - a^{-1} u \\ 0& 1 \end{pmatrix}.
\ea
Taking the determinant of both decompositions and rearranging yields
\ba
\left| a + u c v^\text{T} \right| =
\left| a \right|  
\left| c  \right|  
\left| c^{-1} + v^\text{T} a^{-1}u \right|.
\labeq{wooddet}  
\ea
Inverting both decompositions and equating top-left corners yields
\ba
\left( a + u c v^\text{T} \right)^{-1} 
=
a^{-1}-a^{-1} u 
\left( c^{-1}+ v^\text{T} a^{-1}u \right)^{-1}v^\text{T} a^{-1}.
\labeq{wood}
\ea

\bibliography{concept.bib}

\end{document}